\documentstyle[epsfig]{aipproc}

\begin{document}

\begin{flushright}
  SHEP 97-10 \\
  UM-TH-97-13 \\
  hep-ph/9706215 \\
  2 June 1997 
\end{flushright}

\title{Borel singularities at small $x$}

\author{Kevin D. Anderson$^*$, Douglas A. Ross$^*$ and \\
 Michael G. Sotiropoulos$^{\dagger}$\thanks{Talk presented at the Fifth 
 International Workshop on Deep Inelastic Scattering and QCD, Chicago, 
 IL, 1997}}

\address{$^*$Physics Department, University of Southampton, 
Southampton SO17 1BJ, UK \\
$^{\dagger}$ Randall Laboratory, University of Michigan, 
Ann Arbor, MI 48109, USA}

\maketitle

\begin{abstract}

 D.I.S. at small Bjorken $x$ is considered within the dipole cascade 
 formalism. 
 The running coupling in impact parameter space is introduced in order to 
 parametrize effects that arise from emission of large size dipoles. 
 This results in a new evolution equation for the dipole cascade. 
 Strong coupling effects are analyzed after transforming the evolution 
 equation in Borel ($b$) space. The Borel singularities of the solution 
 are discussed first for the universal part of the dipole cascade and then 
 for the specific process of D.I.S. at small $x$. In the latter case the
 leading infrared renormalon is at $b=1/\beta_0$ indicating the 
 presence of  $1/Q^2$ power corrections for the small-$x$ structure functions. 

\end{abstract}

\section*{Introduction}

 Small-$x$ D.I.S. is a typical example of the class of processes known as 
 semihard processes. They are characterized by the presence of two 
 large scales ordered as \mbox{$\sqrt{s} \gg Q \gg \Lambda_{\mathrm{QCD}}$}. 
 Since the momentum transfer $Q$ involved is large, these processes are 
 amenable to perturbative QCD treatment. 
 The resummation of leading logarithmic corrections of the energy, 
 or $\ln(1/x)$ for D.I.S., is performed by the well known BFKL equation 
 \cite{BFKL}. It is also known that for small enough $x$ 
 the LLA(x) result  obtained \`{a} la  BFKL, although infrared finite, 
 receives contributions from low transverse momentum regions, where 
 observables become sensitive to non-perturbative corrections. 
 It is this region of low transverse momenta that will be considered here. 

 QCD factorization in the small-$x$ regime can be formulated either in 
 transverse momentum space or in impact parameter space through the 
 introduction of the  dipole cascade in the $s$-channel 
 \cite{NikZak1}, \cite{Muell1}.  
 It is the latter formulation that will be used here. Its merit, 
 apart from the simplicity of the final results, is that it organizes the 
 perturbative expansion in terms of sequential soft gluon emissions 
 in the $s$-channel, which in the Coulomb gauge have a clear probabilistic 
 interpretation (to be contrasted with gluon ladders exchanged in the 
 $t$-channel used in the regge description).  
 For fixed $\alpha_s$ this formulation has been shown to be
 equivalent to BFKL. 
 The dipole approach is more suitable for introducing the running coupling 
 because of its radiative nature. Just like in timelike cascades, 
 relevant for jet evolution, the scale in the coupling will be assumed to be 
 the virtuality of the emitted gluon.  
 Once the dipole evolution equation with running coupling is constructed, 
 its solutions are studied in Borel space and their singularities are 
 identified. This analysis concerns the universal part of the dipole cascade 
 which is independent of the external particles involed in the process 
 (i.e. independent of impact factors in the BFKL formalism).
 Finally, for the specific process of D.I.S. at small $x$ it is shown that 
 the leading infrared renormalon (Borel singularity on the positive semiaxis) 
 occurs at $b \beta_0=1$, which indicates power corrections for the 
 structure functions of ${\mathcal O}(\Lambda_{\mathrm QCD}^2/Q^2)$.

\section*{Dipole evolution equation with running coupling}

 Consider small-$x$ D.I.S. in the rest frame of the nucleon target. 
 The QCD factorization theorem for the structure functions takes the form 
 \cite{NikZak1}
 \begin{equation}
 F_{T,L}(x, Q^2) = \frac{Q^2}{4 \pi \alpha_{em}}
 \int_0^1 d z \, \int d^2 {\mathbf r} \, \Phi^{(0)}_{T,L}(z, r) 
 \,\sigma_{d,N}(Y=\ln(z/x), \mathbf{r}) \, .
 \label{struct0}
 \end{equation}
 Here, $\Phi^{(0)}_{T,L}(z,r)$ is the ${\mathcal O}(\alpha_{em})$
 transition probability for $\gamma^\star \rightarrow q \bar{q}$ and
 $\sigma_{dN}(Y, \mathbf{r})$ is the $q$-$\bar{q}$ dipole-nucleon total 
 cross section. The LLA(x) corrections in this formalism are generated by the 
 sequential emission of soft gluons with strictly ordered rapidities. 
 In the large $N_c$ limit the emission of a soft gluon can be thought of as 
 the production of a pair of dipoles each one of which can become the parent 
 for further emissions. The whole cascade can be described in terms of the 
 dipole density $n(Y, r, \rho)$ \cite{MuellPatel} which satisfies the 
 evolution equation 
 \begin{equation}
 \frac{\partial}{\partial Y} n(Y, r, \rho) =
 \int_0^\infty d r^\prime  
 {\mathcal K}(r, r^\prime)  \,
  n(Y, r^\prime, \rho) \, ,
 \, \, \, \, \, \,
 n(Y=0, r, \rho) = r \, \delta(r-\rho)  \, .
 \label{dipev} 
 \end{equation}
 The kernel ${\mathcal K}$ is calculated in perturbation theory to 
 ${\mathcal O}(\alpha_s)$ from the $q \bar{q} \rightarrow q \bar{q} g$ 
 process plus the corresponding virtual non-radiative process. 
 In terms of the dipole density $n$, the cross section $\sigma_{dN}$ becomes 
 \begin{equation}
 \sigma_{dN}(Y, {\mathbf r}) = \int \frac{d^2 \rho}{2 \pi \rho^2} \, 
  n(Y, r, \rho) \, \sigma_0(\rho, m_N) \, .
 \label{sigma_n}
 \end{equation}
 All the large $\ln(1/x)$ corrections are contained in $n$, which is projectile
 and target independent, whereas $\sigma_0$ is the cross section for 
 absorption of dipole of transverse size $\rho$ by the nucleon target. 
 $\sigma_0$ contains information about the nucleon size and it is typically 
 beyond the reach of perturbation theory, unless the nucleon is simulated 
 by a small size onium state. 
 
 The introduction of the running coupling occurs at the level of the basic 
 branching process  $q \bar{q} \rightarrow q \bar{q} g$ and with scale 
 $\alpha_s({\mathbf{k}^2})$, with $\mathbf{k}$ the transverse momentum of the 
 emitted soft gluon. The derivation and the form of the kernel with running 
 coupling are given in ref. \cite{ARS1}. 
 Here it is more fruitful to consider the dipole evolution equation in Borel 
 space. The Borel image $\tilde{n}$ of $n$ with respect to $\alpha_s(Q^2)$ 
 is defined as  
 \begin{equation}
 n(Y, r, \rho; \alpha_s(Q^2) = \int_0^\infty d b \, 
 \tilde{n}(Y, r, \rho; b) \, e^{-b/\alpha_s(Q^2)} \, .
 \label{boreln}
 \end{equation}
 This leads to the following evolution equation in Borel space. 
 \begin{equation}
 \frac{\partial}{\partial Y} \tilde{n}(Y, r, \rho; b) =
 \int_0^\infty d r^\prime
 \int_0^b d b^\prime  
 \tilde{\mathcal K}(r, r^\prime; b^\prime)  \,
 \tilde{n}(Y, r^\prime, \rho; b-b^\prime) \, ,
 \label{dipevol} 
 \end{equation}
 with boundary condition
 \begin{equation}
 \tilde{n}(Y=0, r, \rho; b) = r \, \delta(r-\rho) \, \delta(b) \, .
 \label{bc}
 \end{equation}
 The evolution kernel in Borel space is 
 \begin{eqnarray}
 \tilde{\mathcal K}(r, r^\prime; b) &=& \frac{N_c}{\pi} 
 \left( \frac{Q^2 r^2}{4} \right)^{b \beta_0} 
 \frac{ \Gamma(-b \beta_0)}{\Gamma(1+ b \beta_0)} \delta(r-r^\prime)
 \nonumber \\
 + 2 \frac{N_c}{\pi^2} &\quad& \frac{1}{r^\prime}
 \left( \frac{Q^2 r^{\prime 2}}{4} \right)^{b \beta_0}
 \int_0^1 \frac{ d \omega}{\omega^{1/2} (1-\omega)^{1/2}}
 \frac{ \Gamma(1-\omega b \beta_0) \, \Gamma(1-(1-\omega) b \beta_0)}
      { \Gamma(1+\omega b \beta_0) \, \Gamma(1+(1-\omega) b \beta_0)}
 \nonumber \\
 &\times& 
 \left\{ \left( \frac{ r_>^2}{r^{\prime 2}} \right)^{b \beta_0-1}
  {_2F_1} \left(1- b \beta_0, 1-b \beta_0;1;
                 \frac{r_<^2}{r_>^2}\right) 
 \right.
 \nonumber \\
 &+&  
   \left( \frac{r^2-r^{\prime2}}{ r_>^2} \right)
     \left( \frac{ r_>^2}{r^{\prime 2}} \right)^{\omega b \beta_0}
 {_2F_1} \left(1- \omega b \beta_0, 1-\omega b \beta_0;1;
                  \frac{r_<^2}{r_>^2}\right)
 \nonumber \\
 &-& \left. 
 \left( \frac{r_>^2}{r^{\prime 2}} \right)^{\omega b \beta_0}
 {_2F_1} \left( -\omega b \beta_0, -\omega b \beta_0;1;
                 \frac{r_<^2}{r_>^2}\right)
 \right\} \, ,
 \label{runker}
 \end{eqnarray}
 where $r_< = {\mathrm min}(r, r^\prime)$ and 
 $ r_> = {\mathrm max}(r, r^\prime)$ and 
 $\beta_0=(1/4 \pi) [(11/3) N_c + (2/3) N_f]$.
 The first term in (\ref{runker}) comes from virtual corrections where 
 no new dipole is emitted. The rest comes from real emission. Note that 
 in the infrared limit, where there is emission of dipoles of large size, 
 $r^\prime = r_> \gg r=r_<$, the hypergeometric functions are analytic. 
 This makes $\tilde{K}$ suitable for studying the infrared limit 
 analytically and numerically. 

\section*{Borel singularities and power corrections} 

 Inspecting the kernel (\ref{runker}) it is seen that for fixed parent dipole 
 size $r$ there is a potential singularity at $b=0$. However, using the 
 properties of the hypergeometric functions it can be shown that the $b=0$ 
 singularity is of ultraviolet origin and cancels between the real and the 
 virtual part. Near $b=0$ the kernel is analytic. To exhibit the singularity 
 structure, the kernel is convoluted with test functions $(r^2)^\gamma$. 
 These are known to be eigenfunctions of the kernel in the fixed coupling 
 case. 
 After defining the function $\chi(\gamma, b)$ as
 \begin{equation}
 \int_0^\infty d r^\prime \, 
 \tilde{\mathcal K}(r, r^\prime; b) (r^{\prime 2})^\gamma =
 \frac{N_c}{\pi} \chi(\gamma, b) 
 \left(\frac{Q^2 r^2}{4}\right)^{b \beta_0} (r^2)^{\gamma} \, ,
 \label{chidef}
 \end{equation}
 the expression of $\chi(\gamma, b)$ turns out to be
 \begin{eqnarray}
 \chi(\gamma, b) &=& \frac{\Gamma(-b \beta_0)}{\Gamma(1+ b \beta_0)} 
 \nonumber \\
 &+& \frac{\Gamma(-\gamma-b \beta_0)}{\Gamma(1+ \gamma +b \beta_0)}
 \frac{1}{\pi} 
 \int_0^1 \frac{ d \omega}{\omega^{1/2}{(1-\omega)}^{1/2}}
 \frac{\Gamma(1-\omega b \beta_0)}{\Gamma(1+\omega b \beta_0)}
 \frac{\Gamma(1-(1-\omega) b \beta_0)}{\Gamma(1+(1-\omega) b \beta_0)}
 \nonumber \\
 &\times& \left[ \frac{\Gamma(1+\gamma)}{\Gamma(-\gamma)}
                 \frac{\Gamma(b \beta_0)}{\Gamma(1 - b \beta_0)}
   -2 \frac{\Gamma(1+\gamma +(1-\omega) b \beta_0)}
       {\Gamma(1-\gamma -(1-\omega) b \beta_0)}
 \frac{\Gamma(1+\omega b \beta_0)}{\Gamma(1-\omega b \beta_0)}
 \right] \, ,
 \label{chiform}
 \end{eqnarray}
 where the first term comes from virtual correction and the
 second from real emission.   
 The virtual contribution to $\chi(\gamma, b)$ 
 contains a series of poles at $b \beta_0 = 1, \, 2, \, 3 \, ...$
 which are identified with the IR renormalons
 and correspond to power corrections of ${\mathcal O}( (m_N^2/Q^2)^n)$,
 $ n=1,2,...$
 Note that these poles are independent of the specific form of the
 test function.
 This set of (familiar) poles result from the exponentiation of soft radiation.
 In addition, there is a series of poles at $b \beta_0 = n-\gamma$, 
 $n=0,1,2 ...$ generated by  the $\Gamma(-\gamma-b \beta_0)$ dependence of 
 the real contribution  to $\chi(\gamma, b)$.
 For $Re(\gamma) \ge m$ these poles correspond to IR renormalons for $n>m$.
 Their IR origin is established by observing that these singularities
 arise from the $r^\prime > r$ integration region of eq.~(\ref{dipevol}),
 where the offspring dipole is emitted with size larger than the parent
 dipole.
 Taken at face value the $\gamma$-dependent poles indicate the presence
 of ${\mathcal O}( (m_N^2/Q^2)^{n-\gamma})$ power corrections.
 
 Scale invariance is manifestly broken by the introduction of the running 
 coupling and $(r^2)^\gamma$ are not eigenfuctions of the kernel. 
 It is worth noting though that $\chi(\gamma, b)$ admits power series 
 expansion  around the conformal point $b=0$ of the form 
 \begin{equation}
 \chi(\gamma, b) = \chi(\gamma) +b \beta_0 \chi^{(1)}(\gamma)
  + {\mathcal O}(b^2 \beta_0^2) \, ,
 \label{chiexpand}
 \end{equation}
 \begin{equation}
 \chi^{(1)}(\gamma) =
             -\frac{1}{\gamma} \chi(\gamma)
             -2 \Psi(1) \chi(\gamma) 
             +\frac{1}{2} \chi(\gamma)^2  
             +\frac{1}{2} \chi^\prime(\gamma) \, . 
 \label{chi1}
 \end{equation}
 The ${\mathcal O}(1/b)$ singular terms have cancelled as anticipated
 and the ${\mathcal O}(b^0)$ term is the BFKL spectral
 function $\chi(\gamma)= 2\Psi(1)- \Psi(\gamma)-\Psi(1-\gamma)$.

 Eq. (\ref{dipevol}) is an integral equation in $b$ of the Volterra type, 
 hence for bounded kernel there are no eigenfunctions. However it can 
 be solved formally by iteration. It can then be shown \cite{ARS2} that 
 for fixed $\gamma$ and (say) $ 0 < \gamma < 1$ subsequent iterations 
 of the kernel do not change the position of the leading IR renormalon, 
 which is a branch cut at $b=\gamma/ \beta_0$. 
 This would signal the presence of $\gamma$-dependent power corrections 
 of ${\mathcal O}(m_N^{2 \gamma}/ Q^{2 \gamma})$, 
 and for $\gamma=1/2$, the saddle point in the conformal  (BFKL) limit, 
 this results in $1/Q$ corrections \cite{Levin}. 

 However, as eqs.~(\ref{struct0}) and (\ref{sigma_n}) indicate, 
 the small-$x$ structure  functions are determined by the convolution 
 of the dipole density $n$  with $\Phi^{(0)}$ and $\sigma_0$. 
 Test functions of the form $(r^2)^\gamma$, $\gamma>0$, do not have an 
 IR cutoff for $r \rightarrow \infty$, whereas both $\Phi^{(0)}$ and  
 $\sigma_0$ do. 
 Specifically, convolution with $\sigma_0$ will constrain the emission 
 of arbitrarily large dipoles down to scale $R_N \sim m_N^{-1}$, 
 where $R_N$ is the characteristic length scale or size of the nucleon. 
 Even though $\sigma_0$ cannot be calculated in perturbation theory, to study 
 its effect on the Borel singularities of $\sigma_{dN}$ it is enough 
 to model $\sigma_0(\rho, m_N)$ by a function that regularizes it 
 in the infrared through the scale $R_N$. 
 A simple choice would be to approximate the nucleon by an onium state 
 of size $R_N$. 
 Then, via numerical integration of eq.~(\ref{dipevol}),
 it can be seen  that the leading IR renormalon is a branch cut at 
 $b \beta_0 =1$, signaling the presence of 
 ${\mathcal O} (m_N^2/Q^2)$ power corrections for the structure functions 
 at small $x$ \cite{ARS2}. 
 This is consistent with the expectation from Wilson OPE. 
 We expect this formalism to yield information about the effect of the 
 IR region on the perturbative pomeron intercept.

\end{document}